\documentclass[prl,twocolumn,showpacs,preprintnumbers,amsmath,amssymb]{revtex4}
\usepackage{graphicx}% Include figure files
\usepackage{dcolumn}% Align table columns on decimal point
\usepackage{bm}% bold math

\def\be{\begin{equation}}
\def\ee{\end{equation}}
\def\e#1{\label{#1}\end{equation}}
\def\bea{\begin{eqnarray}}
\def\eea{\end{eqnarray}}
\def\ea#1{\label{#1}\end{eqnarray}}

\def\bem#1{\begin{mathletters}\label{#1}}
\def\eml{\end{mathletters}}

\def\ket#1{{|#1\rangle}}
\def\bra#1{{\langle#1|}}

\def\4#1{{\boldsymbol{#1}}}
\def\8#1{{\widetilde{#1}}}
\def\mean#1{{\langle{#1}\rangle}}

\begin{document}

\preprint{APS/123-QED}

\title{Universal Dephasing Control During Quantum Computation}

\author{Goren Gordon}
\email{goren.gordon@weizmann.ac.il}
\author{Gershon Kurizki}
\email{gershon.kurizki@weizmann.ac.il}
\affiliation{%
Department of Chemical Physics, Weizmann Institute of Science,
Rehovot 76100, Israel
}%

\date{\today}% It is always \today, today,
             %  but any date may be explicitly specified
\begin{abstract}
Dephasing is a ubiquitous phenomenon that leads to the loss of
coherence in quantum systems and the corruption of quantum
information. We present a universal dynamical control approach to
combat dephasing during all stages of quantum computation, namely,
storage, single- and two-qubit operators. We show that (a)
tailoring multi-frequency gate pulses to the dephasing dynamics
can increase fidelity; (b) cross-dephasing, introduced by
entanglement, can be eliminated by appropriate control fields; (c)
counter-intuitively and contrary to previous schemes, one can
increase the gate duration, while simultaneously increasing the
total gate fidelity.
\end{abstract}



\maketitle

Quantum computations, which promise to be faster than their
classical analogs in a range of applications \cite{gro97}, can be
performed via single-qubit and two-qubit operations only
\cite{bar95a}. However, their experimental implementation has been
proven to be difficult due to decoherence effects, which cause
losses of quantum information \cite{los98,scu97}, particularly
dephasing. Moreover, entanglement of qubits via two-qubit gates
\cite{Blatt,two-qubit}, which is the cornerstone of quantum
computation, results in faster loss of computational fidelity due
to dephasing \cite{yu04}.

The problems of a single decohering qubit and its dynamical
control have been thoroughly investigated, and recently extended
to multipartite decoherence control \cite{kof04}. Attempts have
been made to combat dephasing during the storage stage, by
applying sufficiently frequent, fast and strong pulses
\cite{sea00} or by introducing decoherence-free subspaces
\cite{vio00}. However, cross-dephasing due to entanglement, and
the optimization of single- and two-qubit gate pulses so as to
minimize dephasing \cite{los98,hil03,kof04} still need to be
studied.

Here we present a universal dynamical-control approach aimed at
suppressing dephasing during all stages of quantum information
processing, namely, (i) information storage; (ii) manipulation by
single-qubit gates, without entanglement, or (iii) by two-qubit
gates that introduce entanglement. Our main results are to show
that in order to reduce dephasing, it is advantageous to exert
specific, addressable, dynamical control {\em on all the qubits at
once}, whether or not they are manipulated by quantum gates. We
show that the conventional approaches, whereby one tries to either
reduce the gate duration or increase its coherence time, are not
necessarily the best options in our control scheme. Instead, one
can increase the gate duration and simultaneously reduce the
effects of dephasing, resulting in higher gate fidelity. We
introduce the multi-qubit system with its gates' implementation
and arrive at a {\em general solution} to the problem of its
dephasing. We then apply our (analytic) solution to two-qubit
control and its fidelities, and use it in a detailed example of
three-qubit computation.

Our system comprises $N$ qubits, with ground and excited states
$\ket{g}_j, \ket{e}_j$, respectively, and identical excitation
energy $\hbar\omega_0$. Each qubit's excited state experiences
random fluctuations, $\hbar\delta_j(t)$, thus introducing random
dephasing. The total Hamiltonian is given by:
\bea
\label{total-Hamiltonian}
\hat{H}&=&\hat{H}^{(0)}(t)+\hat{H}^{(1)}(t)+\hat{H}^{(2)}(t)\\
\label{zero-Hamiltonian}
\hat{H}^{(0)}(t)&=&\hbar\sum_{j=1}^N\left[\omega_0+\delta_j(t)\right]\ket{e}_{jj}\bra{e}\bigotimes_{k\neq
j}\4I_k\\
\label{single-Hamiltonian}
\hat{H}^{(1)}(t)&=&\hbar\sum_{j=1}^N\left(V^{(1)}_j(t)\ket{e}_{jj}\bra{g}
+ H.c.\right)\bigotimes_{k\neq j}\4I_k\\
\label{two-Hamiltonian}
\hat{H}^{(2)}(t)&=&\hbar\sum_{j=1}^N\sum_{k=j+1}^N
\Big(V^{(2)_\Psi}_{jk}(t)\ket{ge}_{jk}\bra{eg}
\\&&+ V^{(2)_\Phi}_{jk}(t)\ket{ee}_{jk}\bra{gg}+
H.c.\Big)\bigotimes_{l\neq j,k}\4I_l
\eea
where the superscript denotes the manipulation type (e.g., $1$ and
$2$ for one- and two-qubit manipulation, respectively), and the
subscript denotes the subject of manipulation. Here,
$V^{(1)}_j(t)$ is the time-dependent single-qubit gate of the
$j$-th qubit, $V^{(2)_\Psi,(2)_\Phi}_{jk}(t)$ are two possible
time-dependent two-qubit gates, acting on qubits $j$ and $k$,
where the notation is derived from their diagonalization basis,
i.e. the Bell-states basis,
$\ket{\Psi_\pm}=1/\sqrt{2}e^{-i\omega_0t}(\ket{eg}\pm\ket{ge})$,
$\ket{\Phi_\pm}=1/\sqrt{2}(e^{-i2\omega_0t}\ket{ee}\pm\ket{gg})$.
Also, $\4I$ is the identity matrix and $H.c.$ is Hermitian
conjugate.

We treat the random dephasing, differently experienced by each
qubit, as a stochastic Gaussian process with first and second
ensemble-average-moments, $\overline{\delta_j(t)}=0$,
$\Phi_{jk}(t)=\overline{\delta_j(t)\delta_k(0)}$. We assume, for
simplicity, that the driving fields of the single-qubit gates are
resonant, with a time-dependent real envelope, i.e.
$V^{(1)}_j(t)=\Omega^{(1)}_j(t)e^{-i\omega_0t}+c.c.$, and the
driving fields of the two-qubit gates are resonant on their
transition, with a time-dependent real envelope, i.e.
$V^{(2)_\Psi}_{jk}(t)=\Omega^{(2)_\Psi}_{jk}(t)+c.c$, and
$V^{(2)_\Phi}_{jk}(t)=\Omega^{(2)_\Phi}_{jk}(t)e^{-i2\omega_0t}+c.c.$.
The rotating-wave approximation is used.

We consider three generic cases, namely, (a) only single-qubit
gates are applied; (b) only two-qubit gates are applied on
different pairs of qubits; and (c) single- and two-qubit gates are
applied, where each qubit is either manipulated by a single- or a
two-qubit gate, but never by both at once.

The single- ($q=1$) and two- ($q=2$) qubit cases can be solved by
transforming to the interaction picture, and diagonalizing
$\hat{H}^{(q)}(t)$. The diagonalizing basis for the entire
Hamiltonian is then given by $2^N$ basis states,
$\ket{\Psi^{(q)}_{l_q}}=\bigotimes_{j=1}^N\ket{b^{l_q}_j}_j$.
Here, $l_1=0...2^N-1$, $\{b^{l_1}_j\}$ is the binary
representation of $l_1$, meaning
$l_1=b^{l_1}_1b^{l_1}_2...b^{l_1}_N$, with $b^l_j=0,1$
corresponding to
$\ket{\pm}_j=1/\sqrt{2}(e^{-i\omega_0t}\ket{e}_j\pm\ket{g}_j)$,
respectively; and $l_2=0...2^N-1$, $\{c^{l_2}_j\}$ is the quartary
representation of ${l_2}$, meaning
${l_2}=c^{l_2}_1c^{l_2}_2...c^{l_2}_N$, with $c^{l_2}_j=0,1,2,3$
corresponding to $\ket{\Psi_+,\Psi_-,\Phi_+,\Phi_-}_{kk'}$,
respectively. For the density matrix of the ensemble,
$\overline{\rho}(t)=\overline{\ket{\psi}\bra{\psi}}$, where
$\ket{\psi}=\sum_{j=1}^{2^N}\beta_l(t)\ket{\Psi^{(q)}_l}$, the
solution, to second order in $\delta_j(t)$, is then found to be:
\bea
\label{single-solution}
&\overline{\rho}(t)=\rho(0)-\int_0^tdt'\int_0^{t'}dt''
\overline{[\hat{W}^{(q)}(t'),[\hat{W}^{(q)}(t''),\rho(0)]]}\\
\label{single-W}
&\hat{W}^{(q)}(t)=\sum_{l,m=1}^{2^N}w^{(q)}_{lm}(t)\ket{\Psi^{(q)}_l}\bra{\Psi^{(q)}_m}\\
\label{single-w}
&w^{(q)}_{lm}(t)= w^{(q)*}_{ml}(t)=\frac{1}{2}\left\{
\begin{array}{ll}
\delta_j(t)\epsilon^{(1)}_j(t)   & b^{l_1}_j=0, b^{m_1}_j=1  \\& b^{l_1}_k=b^{m_1}_k \,\,\forall k\neq j\\
\delta_{j-}\epsilon^{(2)_\Psi}_j & b^{l_2}_j=0, b^{m_2}_j=1  \\& b^{l_2}_k=c^{m_2}_k \,\,\forall k\neq j\\
\delta_{j+}\epsilon^{(2)_\Phi}_j & b^{l_2}_j=2, b^{m_2}_j=3  \\& b^{l_2}_k=c^{m_2}_k \,\,\forall k\neq j\\
  0 & {\rm otherwise}
\end{array}
\right.
\eea
where $\rho(0)$ is the initial density matrix, given in the
diagonalized basis,
$\delta_{j\pm}(t)=\delta_k(t)\pm\delta_{k'}(t)$. Here,
$\epsilon^{(1)}_j(t)=e^{i\phi^{(1)}_j(t)}$,
$\epsilon^{(2_\Psi,2_\Phi)}_j(t)=e^{i\phi^{(2_\Psi,2_\Phi)}_{p_j}(t)}$,
where $\phi^{(1)}_j(t)=\int_0^tdt'\Omega^{(1)}_j(t')$ and
$\phi^{(2_\Psi,2_\Phi)}_{p_j}(t)=\int_0^tdt'\Omega^{(2_\Psi,2_\Phi)}_{p_j}(t')$
are the accumulated phases.

This is the most general scheme of dephasing control analyzed thus
far, in that it satisfies all the requirements for quantum
computation, namely single- and two-qubit gates, applied
simultaneously on different qubits. It is solvable by combining
the results above with the solution given in the form of
Eqs.~\eqref{single-solution}, where the general interaction
operator $\hat{W}(t)$ is a combination of Eqs.~\eqref{single-w}
with $q=1,2$. The three stages of quantum computation are defined
by the restrictions on the overall phase accumulated by the state,
due to the application of the gate fields at the end of each
stage. During storage, the restrictions are $\phi^{(q)}_j(T)=2\pi
M_j\quad M_j=0,\pm1,\ldots$. To implement a Hadamard gate applied
to the $j$-th qubit \cite{nie00}, the restrictions are
$\phi^{(1)}_j(T)=\pi/4$ and storage regime for all the rest. To
implement a SWAP gate between qubits $k$ and $k'$ \cite{los98},
the restrictions are $\phi^{(2)_\Psi}_{kk'}(T)=\pi/4$ and storage
regime for all the rest.

To characterize the efficiency of the dephasing control schemes,
we use fidelity, defined as $F(T)={\rm Tr}(\rho_{\rm
target}^{1/2}\overline{\rho}(T)\rho_{\rm target}^{1/2})$, where
$\rho_{\rm target}$ is the target density matrix after the quantum
computation, e.g. $\rho^{\rm target}=\rho(0)$ for the storage
stage. The error of the gate operation is then $E(T)=1-F(T)$.
However, since quantum computations require lack of knowledge of
the initial qubits' state, we shall also use the average fidelity,
$F_{avg}(T) = \mean{F(T)}$, where $\mean{\cdots}$ is the average
over all possible initial pure states.

Armed with the general solutions and efficiency measures presented
above, we now analyze in detail quantum computation by two qubits
experiencing random dephasing. First, we apply single-qubit gates
on each of the qubits. The average fidelity of this scheme is
given by:
\bea
\label{two-qubits-avg-fidelity-1}
&F_{avg}(T)=1-\frac{5}{12}\left(J^{(1)}_{11}(T)+J^{(1)}_{22}(T)\right)\\
\label{J-1-def}
&J^{(q)}_{jk}(t) =
\int_0^tdt'\int_0^{t'}dt''\Phi_{jk}(t'-t'')\epsilon^{(q)}_j(t')\epsilon^{(q)*}_k(t'')\\
\label{J-1-def-f-domain}
&{\rm Re}J^{(q)}_{jk}(t) = \pi\int_{-\infty}^\infty d\omega
G_{jk}(\omega)\epsilon^{(q)}_{j,t}(\omega)\epsilon^{(q)*}_{k,t}(\omega)
\eea
where $J^{(q)}_{jk}(t)$ is the modified dephasing function due to
fields $\Omega^{(q)}_{j,k}$, $q=1,2_\Psi,2_\Phi$. Here,
$G_{jk}(\omega)=(2\pi)^{-1}\int_{-\infty}^\infty
dt\Phi_{jk}(t)e^{i\omega t}$ is the dephasing spectrum, and
$\epsilon^{(q)}_{j,t}(\omega)=(2\pi)^{-1/2}\int_0^t
dt'\epsilon^{(q)}_j(t')e^{i\omega t'}$ is the finite-time Fourier
transform of the modulation (for a thorough analysis of the
modified dephasing function, see Refs.~\cite{kof04}).

Equations~\eqref{two-qubits-avg-fidelity-1}-\eqref{J-1-def-f-domain}
show the dependence of the fidelity on the spectral
characteristics of the fields and the dephasing, and suggest how
to tailor specific gate and control pulses: {\em reducing the
spectral overlap} of the dephasing and modulation spectra
\cite{kof04}. Furthermore, they show that single-qubit gate fields
do not cause cross-dephasing, since
Eq.~\eqref{two-qubits-avg-fidelity-1} depends only on single-qubit
dephasing, $\Phi_{jj}(t)$. This comes about from the averaging
over all initial qubits, where for each initial entangled state
that ``suffers'' from cross-dephasing (e.g. triplet,
$\ket{\Phi_-}$), there is another entangled state that
``benefits'' from cross-dephasing (e.g. the singlet,
$\ket{\Psi_-}$). Equation~\eqref{two-qubits-avg-fidelity-1} also
shows that the modified dephasing function appears no matter what
the accumulated phase, meaning that if one applies a gate field on
one qubit, one can still benefit from applying a control field on
the other, stored, qubit.

Next, we explore two-qubit gate operations. The average fidelity
for this task is found to be:
\be
\label{two-qubit-avg-fidelity-2}
F_{avg}(T)=1-\frac{5}{24}\sum_{j,k=1,2}\left(J^{(2)_\Phi}_{jk}(T)+
(-1)^{j+k}J^{(2)_\Psi}_{jk}(T)\right)
\ee
Here we see that cross-dephasing does not cancel due to averaging,
but has opposite signs for the different two-gate fields. Thus,
for example, a SWAP gate may benefit from cross-dephasing.
Furthermore, we see that applying both two-qubit gate fields can
reduce dephasing, even if only one field is needed for the actual
gate operation. This means that applying a two-qubit storage gate
field, with $\phi^{(2)_\Phi}_{1,2}(T)=2\pi M$, $M=1,2,\ldots$,
along with, e.g., a SWAP gate, can reduce dephasing.

This novel approach, consisting in applying a storage gate field,
concurrently with the actual gate field required for the logic
operation, may result in longer gate durations, due to limitations
on the gate fields themselves, such as minimal duration and
maximal achievable peak-power. However, as seen from
Eqs.~\eqref{two-qubits-avg-fidelity-1},~\eqref{two-qubit-avg-fidelity-2},
this may still be beneficial if the reduction in the modulated
dephasing due to the applied fields is greater than its increase
due to longer gate duration.

\begin{figure}[htb]
\centering\includegraphics[width=8.5cm]{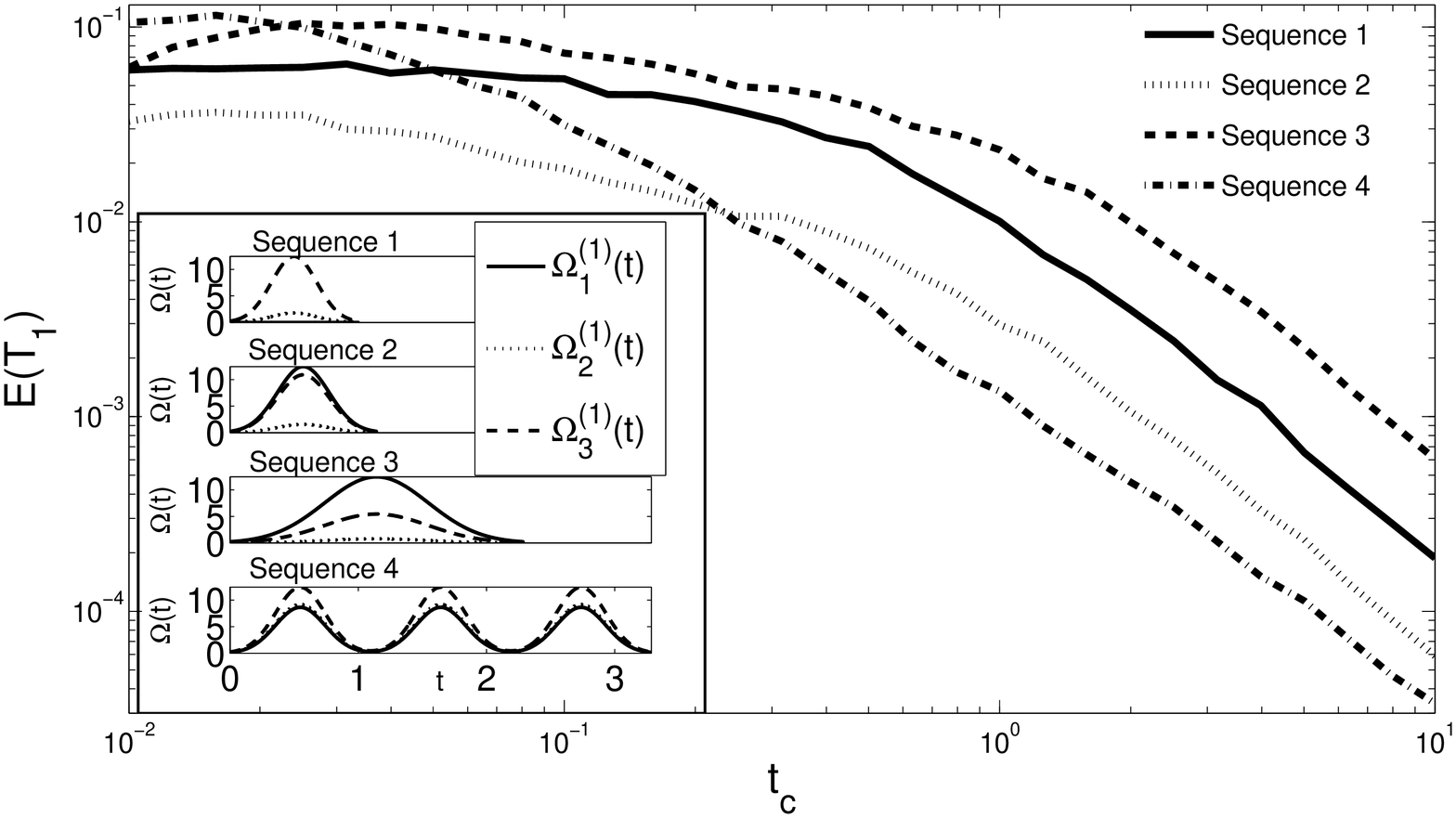}
\protect\caption{Gate errors at the end of the first stage,
$E(T_1)$ as a function of correlation time, $t_c$, for different
dynamical gate fields [inset]. The gate fields parameters are
$\phi^{(1)}_{1,2,3}(T_1)=\{0,\pi/4,7\pi/4\}$,
$\{2\pi,\pi/4,7\pi/4\}$, $\{4\pi,\pi/4,7\pi/4\}$ and
$\{4\pi,17\pi/4,23\pi/4\}$ for sequences $1-4$, respectively. The
gate durations are chosen such that the peak-power is the same for
all sequences. Here $\gamma=0.1$, and the results were obtained
after averaging over 1000 realizations.}
\label{fig-first-stage}\end{figure}

We explore this approach in a specific, complex scenario of three
qubits experiencing random dephasing. We accompany it by numerical
results, where the dephasing is taken to be a set of random
fluctuations with correlation-functions
$\Phi_{jk}(t)=(\gamma/t_{c,jk})e^{-t/t_{c,jk}}\xi(|r_{jk}|)$,
where $\gamma$ is the asymptotic dephasing rate (without control
fields $J(t\gg t_c)=\gamma t$), $t_{c,jk}$ is the corresponding
correlation-time, and $\xi(|r_{jk}|)$ is the qubit-distance
dependent cross-dephasing overlap function, with $\xi=0(1)$
denoting no (maximal) cross-dephasing. We shall take
$t_{c,jk}=t_c$, $\forall j,k$. For the gate fields we take
realistic Gaussian pulses, with a restriction on their minimal
duration and maximal peak-power,
Fig.~\ref{fig-first-stage}[inset]. The additional restrictions on
the gate phases, e.g. $\phi^{(q)}_j(T)=2\pi M_j$, leaves only one
free parameter per field, namely $M_j$.

The initial state of the system is picked to be
$\ket{\psi(0)}=\ket{\uparrow}_1\ket{e}_2\ket{\downarrow}_3$, where
$\ket{\uparrow(\downarrow)}=\frac{1}{\sqrt{2}}\left(i\ket{1}\pm\ket{0}\right)$.
Three single-qubit gates are applied, one per qubit, for time
$T_1$. Then we store the first qubit, and apply gates on the other
two, with the restrictions on the gate pulses being
$\phi^{(1)}_{1}(T_1)=2\pi M$; $\phi^{(1)}_{2}(T_1)=\pi/4$; and
$\phi^{(1)}_{3}(T_1)=7\pi/4$. The desired (target) state at the
end of this stage is $\psi^{\rm
target}(T_1)=i\ket{\uparrow}_1\ket{\uparrow}_2\ket{g}_3$.
In order to demonstrate the advantageous effects of complex
dynamical gate sequences, and the benefits of longer gate
durations while controlling the stored qubit, we compare our
proposed approach (Fig.~\ref{fig-first-stage}[inset]) and the
conventional approach, whereby maximal peak-power and minimal
duration Gaussian pulses are applied to achieve the required gates
(Fig.~\ref{fig-first-stage}, sequence $1$). Our proposed approach
requires longer gates, as the Gaussian pulses have a limited
peak-power, and minimal duration. Comparing the proposed sequences
$2$ and $3$ to the conventional sequence $1$, demonstrates the
trade-off between the beneficial effects of controlling the stored
qubit and the detrimental effects of longer duration. For long
correlation times, which are present in several experimental
setups \cite{hu06}, additional increase in fidelity, shown in
sequence $4$, can be achieved by trains of short pulses, which
reduce the dephasing due to higher frequencies in the dynamical
control fields \cite{kof04}, in spite of its more than three-fold
gate duration.

The next stage is to apply a two-qubit gate between the second and
third qubits, and store the first one. We assume that
$\rho(0)=\ket{\psi^{\rm target}(T_1)}\bra{\psi^{\rm
target}(T_1)}$, apply the $\Omega^{(2)_\Psi}_{23}$ two-qubit gate
on the second and third qubits, and use the other two gates to
control the dephasing, resulting in gate pulse restrictions,
$\phi^{(1)}_{1}(T_2)=2\pi M$, $\phi^{(2)_\Psi}_{23}(T_2)=3\pi/2$,
$\phi^{(2)_\Phi}_{23}(T_2)=2\pi M$. The desired state at the end
of this stage is $\ket{\psi^{\rm
target}(T_1+T_2)}=-i\ket{\uparrow}_1\ket{g}_2\ket{-}$.
\begin{figure}[htb]
\centering\includegraphics[width=8.5cm]{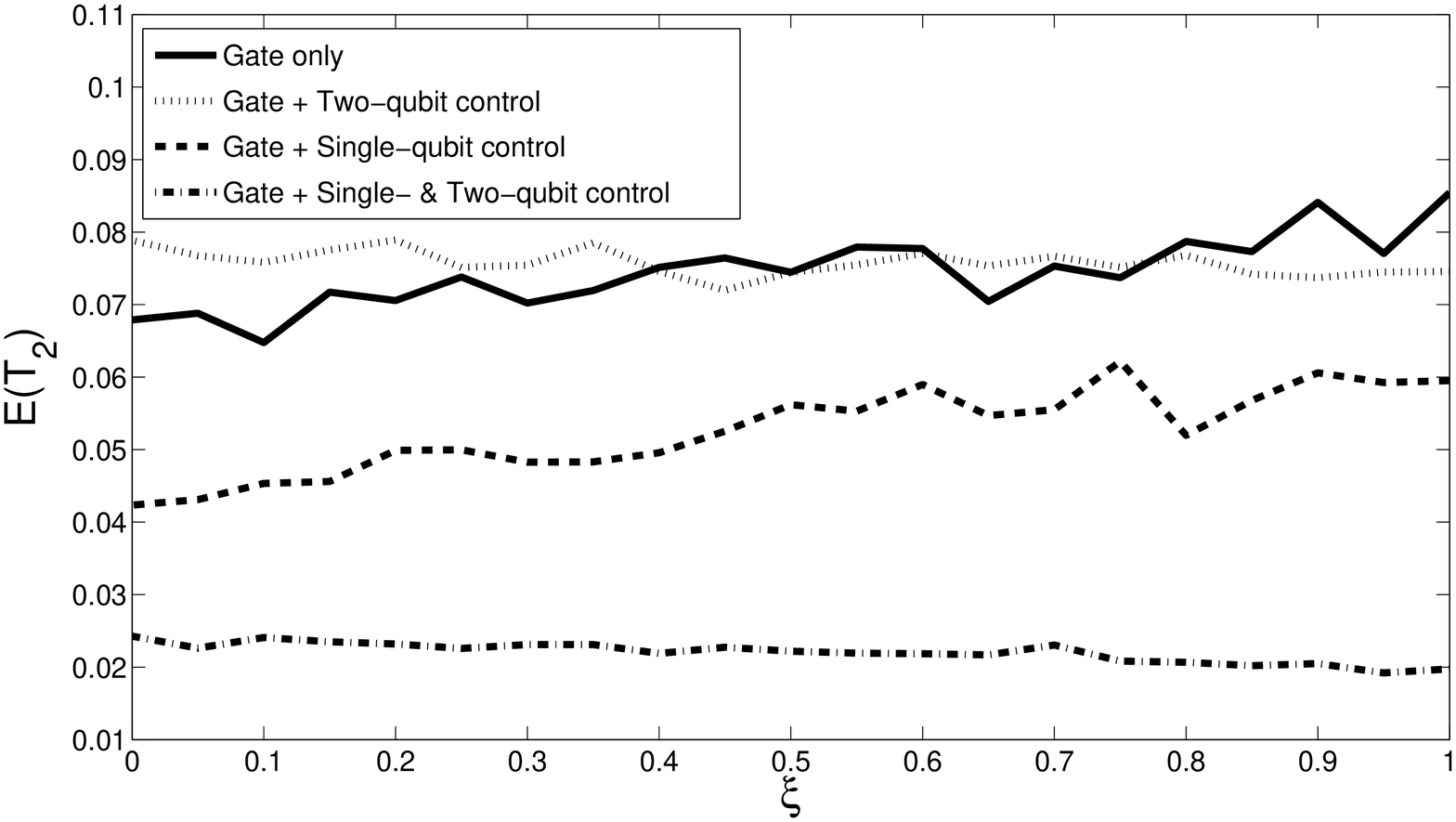}
\protect\caption{Gate errors at the end of the second stage,
$E(T_2)$ as a function of cross-dephasing overlap, $\xi$, for
different dynamical gate fields. The gate fields parameters are
$\{\phi^{(1)}_{1},\phi^{(2)_\Psi}_{23},\phi^{(2)_\Phi}_{23}\} =$
$\{0,3\pi/2,0\}$ (solid), $\{0,3\pi/2,2\pi\}$ (dotted),
$\{2\pi,3\pi/2,0\}$ (dashed) and $\{2\pi,3\pi/2,2\pi\}$
(dash-dot). The gate durations are chosen such that the peak-power
is the same for all sequences. Here $\gamma=0.1$, and the results
were obtained after averaging over 1000 realizations.}
\label{fig-second-stage}\end{figure}
Figure~\ref{fig-second-stage} illustrates the effects of adding
control fields concurrently with the desired SWAP gate. The
application of the gate field reduces the modified dephasing
function, $J^{(2)_\Psi}_{23}$, making the other functions more
dominant. Thus, one can observe the cross-dephasing
overlap-dependent increase in gate error. However, introducing a
control field for the second two-qubit gate eliminates this
cross-dephasing dependence, resulting in a trade-off between
gate-field duration increase and cross-dephasing decrease.
Furthermore, introducing only control of the single-qubit field
reduces the error, but leaves the cross-dephasing intact.
Combining the two control schemes results in an even greater
decrease in error, without cross-dephasing.
\begin{figure}[htb]
\centering\includegraphics[width=8.5cm]{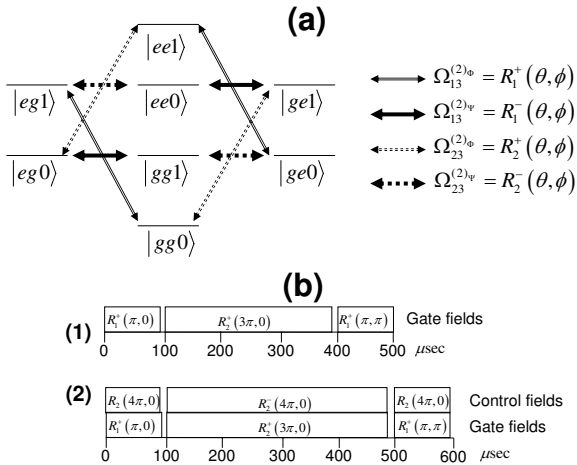}
\protect\caption{(a) Schematic diagram of energy levels and
two-qubit gate fields applied for two internal states of two ions
$\ket{g(e)}$ and first two common vibrational levels $\ket{0(1)}$.
(b) Conventional pulse sequence (1) and our proposed sequence (2).
The pulse notation and parameters are taken from
Ref.~\cite{Blatt}.}
\label{fig-ion-trap}\end{figure}

One implementation of these schemes may involve a string of ions
in a linear trap \cite{cir95,Blatt}. The qubits are encoded by two
internal states of each ion ($\ket{g(e)}_j$), and are manipulated
by individual-addressing laser beams. One introduces another
qubit, encoded by the ground and first excited common vibrational
levels ('bus-mode', $\ket{0(1)}_N$). The qubit gates are realized
by applying laser pulses on the 'carrier' ($\Omega_j^{(1)}(t)$,
$\ket{g}\leftrightarrow\ket{e}$), 'blue-sideband'
($\Omega_{jN}^{(2)_\Phi}(t)$,
$\ket{g}\ket{0}\leftrightarrow\ket{e}\ket{1}$) and 'red-sideband'
($\Omega_{jN}^{(2)_\Psi}(t)$,
$\ket{g}\ket{1}\leftrightarrow\ket{e}\ket{0}$) of the electronic
quadrupole transition, Fig.~\ref{fig-ion-trap}(a). However, in a
harmonic trap \cite{Blatt}, the blue- (red-) sideband also couples
to higher excitation levels, e.g.
$\ket{g}\ket{1}\leftrightarrow\ket{e}\ket{2}$
($\ket{e}\ket{1}\leftrightarrow\ket{g}\ket{2}$) and thus
complicates the concurrent application of both two-qubit gates.
This can be circumvented by imposing anharmonicity on the linear
trap. Dephasing in the ion trap system can appear due to ambient
magnetic field fluctuations that cause a Zeeman shift in the qubit
levels. We have simulated a SWAP gate of two ions, using the first
two common vibrational levels (assuming anharmonicity), with
dephasing ($\gamma^{-1}=1m{\rm sec}$, $t_c=300\mu {\rm sec}$),
Fig.~\ref{fig-ion-trap}(b). We compared the conventional pulse
sequence Fig.~\ref{fig-ion-trap}(b.1), resulting in
$F_{avg}(t=500\mu {\rm sec})=0.93$ and our proposed sequence
Fig.~\ref{fig-ion-trap}(b.2), resulting in $F_{avg}(t=600\mu {\rm
sec})=0.97$. This shows a considerable improvement in gate
fidelity, despite its longer duration.

To conclude, we have formulated a universal protocol for dynamical
dephasing control during all stages of quantum information
processing, namely, storage, single- and two-qubit gate
operations. It amounts to controlling all the qubits, whether they
participate in the computation or not, and tailoring specific gate
and control fields that optimally reduce the dephasing. This
counter-intuitive protocol has a great advantage over others in
that it increases the fidelity of the operation required, whether
storage, manipulation or computation, despite the fact that it
requires longer duration.

We acknowledge the support of GIF and EC (SCALA IP).

\end{document}